# Incorporating AI Incident Reporting into Telecommunications Law and Policy: Insights from India

Avinash Agarwal[a,1], Manisha J. Nene[b]

[a]*Department of Telecommunications, Ministry of Communications, Sanchar Bhawan, New Delhi, 110001, Delhi, India* [b]*Defence Institute of Advanced Technology, Ministry of Defence, Girinagar, Pune, 411025, Maharashtra, India*

## Abstract

The integration of artificial intelligence (AI) into telecommunications infrastructure introduces novel risks, such as algorithmic bias and unpredictable system behavior, that fall outside the scope of traditional cybersecurity and data protection frameworks. This paper introduces a precise definition and a detailed typology of *telecommunications AI incidents*, establishing them as a distinct category of risk that extends beyond conventional cybersecurity and data protection breaches. It argues for their recognition as a distinct regulatory concern. Using India as a case study for jurisdictions that lack a horizontal AI law, the paper analyzes the country's key digital regulations. The analysis reveals that India's existing legal instruments, including the Telecommunications Act, 2023, the CERT-In Rules, and the Digital Personal Data Protection Act, 2023, focus on cybersecurity and data breaches, creating a significant regulatory gap for AI-specific operational incidents, such as performance degradation and algorithmic bias. The paper also examines structural barriers to disclosure and the limitations of existing AI incident repositories. Based on these findings, the paper proposes targeted policy recommendations centered on integrating AI incident reporting into India's existing telecom governance. Key proposals include mandating reporting for high-risk AI failures, designating an existing government body as a nodal agency to manage incident data, and developing standardized reporting frameworks. These recommendations aim to enhance regulatory clarity and strengthen long-term resilience, offering a pragmatic and replicable blueprint for other nations seeking to govern AI risks within their existing sectoral frameworks.

*Keywords:* Telecommunications AI incidents, AI incident reporting, Sectoral regulation, AI governance, Telecom regulations, India, Regulatory gaps, Telecommunications resilience

## 1. Introduction

The rapid adoption of Artificial Intelligence (AI) in telecommunications networks and services has brought transformative improvements in efficiency, scalability, and automation. AI is now integral to critical functions such as network traffic management, adaptive spectrum reuse, fault tolerance, and beamforming [1]. However, its growing role also introduces new risks due to autonomous learning, unpredictability, and systemic interactions [2]. These risks, including biased algorithms [3], unintended model behavior such as model drift [4], and adversarial attacks, can jeopardize the resilience of telecommunications infrastructure, particularly in its critical components. Unlike traditional cybersecurity or data protection incidents, these AI-specific failures present unique challenges for which broader regulations and use cases are yet to mature [5], creating a regulatory gap that this paper explores through a focused case study of India. The problem is compounded by the absence of a specific regulatory mandate for reporting AI incidents [6, 7], making it difficult to assess systemic vulnerabilities and enhance resilience.

To ground this analysis in a concrete technical context, consider a prime example of these critical functions: the use of AI in 5G Radio Access Network (RAN) management. These core infrastructure systems perform tasks like dynamic resource allocation and beamforming. They are not consumer-facing and make autonomous operational decisions, creating subtle and hard-to-detect risks of bias. For example, an AI model for beamforming might be trained on data collected unevenly from its intended operational area, creating a "spatial bias." As the O-RAN Alliance highlights, if the model is trained primarily on data from users along a specific diagonal path in a sample square area, and then validated using a test set from that same biased data, it will appear to function perfectly. However, in a real-world deployment, it could systematically provide suboptimal or degraded service to users in areas other than the diagonal path. This is a flaw that would be invisible to the operator's standard testing procedures [8]. Further, AI bias in these core systems can inadvertently automate and amplify existing inequities.

[1] Corresponding author

*Email addresses:* avinash.70@gov.in (Avinash Agarwal), mjnene@diat.ac.in (Manisha J. Nene)

[1.] The views expressed here are personal and do not necessarily reflect those of any organisation or institution.

Ericsson has noted that historical investment patterns often lead to better network infrastructure and different usage data in certain areas. An AI model trained on this skewed data may then learn to operate networks that ”work well only in those regions,” systematically providing a lower quality of service to less represented, often lower-income or rural, user groups [9]. It is this class of complex, AI-specific failures within critical infrastructure that motivates the need for a new regulatory perspective: incidents that perpetuate bias or inefficiency, are not security breaches, and can evade conventional quality assurance.

It is acknowledged that some jurisdictions, notably the European Union with its comprehensive AI Act [10], are creating horizontal regulations that establish broad reporting obligations for high-risk AI systems. However, the vast majority of nations have not yet adopted this model. This paper, therefore, addresses the critical and more common scenario where overarching AI governance is absent, examining India as a case study. It argues that for such countries, the most effective and pragmatic approach to enhancing telecom resilience is to integrate AI-specific incident reporting requirements directly into their existing, sector-specific regulatory frameworks.

India is selected as a case study for several compelling reasons. First, as the world's most populous nation [11] and second-largest telecommunications market with over a billion subscribers [12], the scale of AI deployment and the potential impact of any incidents are globally significant. Second, India is undergoing one of the world's fastest 5G rollouts, meaning advanced AI-driven network management systems are being deployed at an unprecedented pace [13, 14]. Third, its regulatory environment is rapidly evolving; with a new Telecommunications Act, 2023 [15] and Digital Personal Data Protection Act, 2023 [16] recently enacted and their implementing rules currently being framed, this analysis offers timely insights for active policy development. Finally, India, like most nations, does not yet have a horizontal, cross-sector AI law, making it a perfect archetype for examining the vital role of sector-specific regulation in managing AI risks. The findings and recommendations are therefore intended to provide a replicable model for other jurisdictions facing similar regulatory circumstances.

Resilience, defined as the ability to anticipate, withstand, and recover from disruptions [17], has become an urgent priority in telecommunications policy. As global digital ecosystems expand, threats to resilience, whether from climate events [18], cyberattacks [19], or geopolitical conflicts [20], cross national borders and impact interconnected networks worldwide. In this context, AI incidents in critical telecom infrastructure pose significant risks, ranging from service disruptions to breaches in trust and accountability [21]. Despite the increasing deployment of AI in telecommunications, as this study shows, existing regulations focus primarily on cybersecurity and data protection but lack provisions to systematically document, analyze, and address AI-specific failures.

The significance of these regulatory gaps is underscored by the specific risks that AI-powered systems introduce to the global telecommunications industry. Adversaries could exploit AI-driven telecom infrastructure to disrupt services, compromise data, or manipulate decision-making [22]. Vulnerabilities in AI-based telecom equipment, especially in a vendor-dominated market, risk large-scale outages with global repercussions. While not AI-related, the Microsoft-CrowdStrike outage in July 2024, which disrupted airport operations worldwide [23], underscores such broader risks. Geopolitical tensions further heighten these threats, as AI systems may be exploited to disrupt services or compromise security. Additionally, the lack of reliable methods to ensure trust in AI decisions limits its use in mission-critical contexts, highlighting the need for more interpretable and resilient AI systems [24]. A key missing element in current regulatory frameworks is AI incident data collection, which relies solely on private initiatives with no clear accountability for data entry, categorization, or oversight [25, 26].

Incident reporting plays a crucial role in enhancing resilience across critical sectors. Established safety-critical domains such as aviation, cybersecurity, and fire safety systematically document incidents to improve safety and prevent recurring issues [21, 27]. In contrast, AI systems often lack a structured approach to learning from past failures, leading to repeated design and deployment mistakes [28]. In telecommunications, AI incident reporting could similarly provide insights into systemic vulnerabilities, enabling proactive measures to prevent disruptions. Such a framework would allow stakeholders to learn from incidents, develop mitigation strategies, and enhance trust in telecom networks and services. Analyzing AI incident repositories is potentially valuable for developers and policymakers and can aid in AI regulation formulation [25]. Yet, a large category of AI incidents remains uncatalogued, unlike cybersecurity or data protection breaches, which have established regulatory mechanisms. The lack of comprehensive incident data hampers efforts to identify patterns, assess risks, and implement effective policy interventions [6]. Addressing this gap is crucial for strengthening the resilience of telecommunications networks increasingly reliant on AI.

Recognizing the need for global coordination, the International Telecommunication Union (ITU) has encouraged the establishment of National Computer Incident Response Teams (CIRTs) and cross-border information sharing through ITU-D Resolution 69 [29] and ITU-T Resolutions 50 [30] and 58 [31]. However, these efforts primarily address cybersecurity threats, leaving other AI-specific and data protection risks unaddressed.

A clear definition of *telecommunications AI incidents* is vital for creating a standardized reporting framework.

This paper proposes defining these incidents as events involving AI systems in critical telecommunications infrastructure that result in disruption, degradation, or manipulation of services; unauthorized access to or misuse of network resources; or harm to individuals or the environment. The definition proposed in this paper expands the scope of AI incidents beyond traditional cybersecurity or data protection events.

Legislative frameworks play a crucial role in shaping telecom network development, particularly for high-risk applications, even though diverse global regulations complicate implementation [32]. Updating telecom regulations to address AI-related risks is essential for enhancing resilience. This shift requires moving from reactive policies to proactive measures that account for the evolving nature of AI, as seen in approaches like dynamic regulation and innovation ecosystems, which promote responsible AI innovation and investment [33]. Regulatory interventions should prioritize mandatory AI incident reporting, incentivizing information sharing while ensuring data anonymization [7].

While several general-purpose AI incident repositories exist, such as the AI Incident Database (AIID) [34] and AIAAIC [35], these lack the granularity required for addressing the complexities of telecommunications networks [36]. Such databases often focus on broad AI use cases, neglecting sector-specific issues critical to telecom resilience. Moreover, these repositories lack regulatory mandates resulting in limited adoption and underrepresentation of incidents [7]. In telecommunications, reporting mechanisms for cybersecurity and data protection breaches are well-established, but they do not extend to other AI-specific incidents, as discussed in this paper. This disconnect leaves operators and regulators without the necessary tools to identify, assess, and mitigate AI-related risks effectively. Addressing this limitation requires a regulatory framework that mandates AI incident reporting tailored to the specific needs of critical telecom infrastructure.

This work aims to contribute to the broader agenda of resilient telecommunications policies by addressing the regulatory gaps in AI incident management. By emphasizing the importance of reporting mechanisms and targeted regulatory interventions, this paper seeks to lay the groundwork for policies that enhance trust, safety, and resilience in critical telecommunications infrastructure.

This paper makes three key contributions toward building resilient telecommunications policies:

1. *Definition and Typology of Telecommunications AI Incidents*: The paper introduces a precise definition of "telecommunications AI incidents," establishing them as a distinct category of risk and thereby expanding the scope of reportable events beyond traditional cybersecurity and data protection violations, providing the legal foundation for a more comprehensive regulatory approach. Crucially, it supports this definition with a detailed typology of plausible incidents, grounded in operational realities, that clarifies the nature of risks falling outside conventional regulatory frameworks. This conceptual clarity provides the foundation for identifying AI-specific risks in telecommunications networks and services.
2. *Identification of Regulatory Gaps in AI Risk Management for Critical Telecommunications Infrastructure*: The paper analyzes India's existing regulatory landscape to identify gaps in addressing AI related risks, particularly those not directly related to cyberattacks or data breaches. It highlights the need for AI incident reporting as a means to enhance resilience and trustworthiness in telecommunications operations.
3. *Actionable Policy Recommendations for AI Incident Reporting in Critical Telecom Networks*: The paper provides eight practical policy recommendations, including mandatory AI incident reporting, standardized reporting frameworks, and the designation of a nodal agency to oversee the collection and management of AI incident data within telecommunications infrastructure.

The rest of this paper is structured as follows: Section 2 presents the research questions and methodology. Section 3 defines *telecommunications AI incidents*, provides illustrative examples and examines how such incidents overlap with cybersecurity incidents and data protection violations. Section 4 analyzes the Indian regulatory framework to assess its adequacy in addressing AI-related risks. Section 5 examines barriers to AI incident reporting and identifies gaps in existing reporting mechanisms relevant to telecommunications. Section 6 proposes eight regulatory and policy measures to bridge these gaps. Section 7 discusses the implications of these findings for telecommunications resilience and suggests directions for future research. Finally, Section 8 concludes with a summary of the key contributions and the likely impact of this work.

## 2. Research Questions and Methodology

This study follows a systematic approach to investigate the integration of AI incident reporting into telecommunications policy, with a focus on resilience.

### *2.1. Research Questions*

The methodology is designed to answer the following four research questions (RQs):

1. *RQ1: How do AI incidents in critical telecommunications infrastructure relate to traditional cybersecurity and data protection incidents, and why is their reporting crucial for resilience?*
2. *RQ2: To what extent does the regulatory framework of India address AI-related risks in critical*

*telecommunications infrastructure alongside cybersecurity and data protection threats?*

3. *RQ3: What are the gaps, barriers, and challenges in current AI incident reporting mechanisms concerning telecommunications resilience?*
4. *RQ4: What regulatory and policy measures can enhance AI incident reporting within telecommunications frameworks?*

### *2.2. Methodology*

This study employs a qualitative case study methodology focused on India. While the analysis is specific, the case was chosen for its archetypal nature, allowing for the generation of insights relevant to other nations with similar regulatory profiles. The methodology comprises the following steps:

1. *Define and Contextualize Telecommunications AI Incidents:* Establish a precise definition of telecommunications AI incidents and contextualize it in relation to traditional cybersecurity and data protection incidents.
2. *Analyze the Indian Regulatory Framework:* Analyze India's telecommunications, cybersecurity, and data protection regulations to assess the extent of AI incident reporting coverage and identify specific legal and policy gaps that hinder effective AI risk management in telecommunications.
3. *Analyze Shortcomings and Barriers in AI Incident Reporting:* Identify barriers to AI incident reporting and evaluate the absence or inadequacy of existing AI incident repositories, particularly in critical infrastructure.
4. *Propose Policy Recommendations:* Develop regulatory interventions tailored to the Indian context to integrate AI incident reporting into telecommunications governance. The recommendations address the four identified regulatory gaps and barriers to reporting while considering implementation challenges and balancing regulatory oversight with industry feasibility.

Figure 1 shows an overview of the research methodology followed in this paper. This structured approach was designed to systematically address the research questions by identifying gaps in India's existing framework and facilitating data-driven policy-making and research to improve resilience in telecommunications.

## 3. Defining Telecommunications AI Incidents and their relation to Cybersecurity and Data Protection

Before analyzing the specific gaps in India's regulatory framework, it is essential to establish a clear conceptual foundation. This section proposes a precise, domain-specific definition of a *Telecommunications AI Incident*. This definition is not presented as a foregone conclusion, but rather as an analytical tool that enables a more nuanced examination of existing laws. By defining the full spectrum of potential AI-induced harms, many of which are not malicious security breaches, we can more effectively assess whether current regulations are fit for purpose.

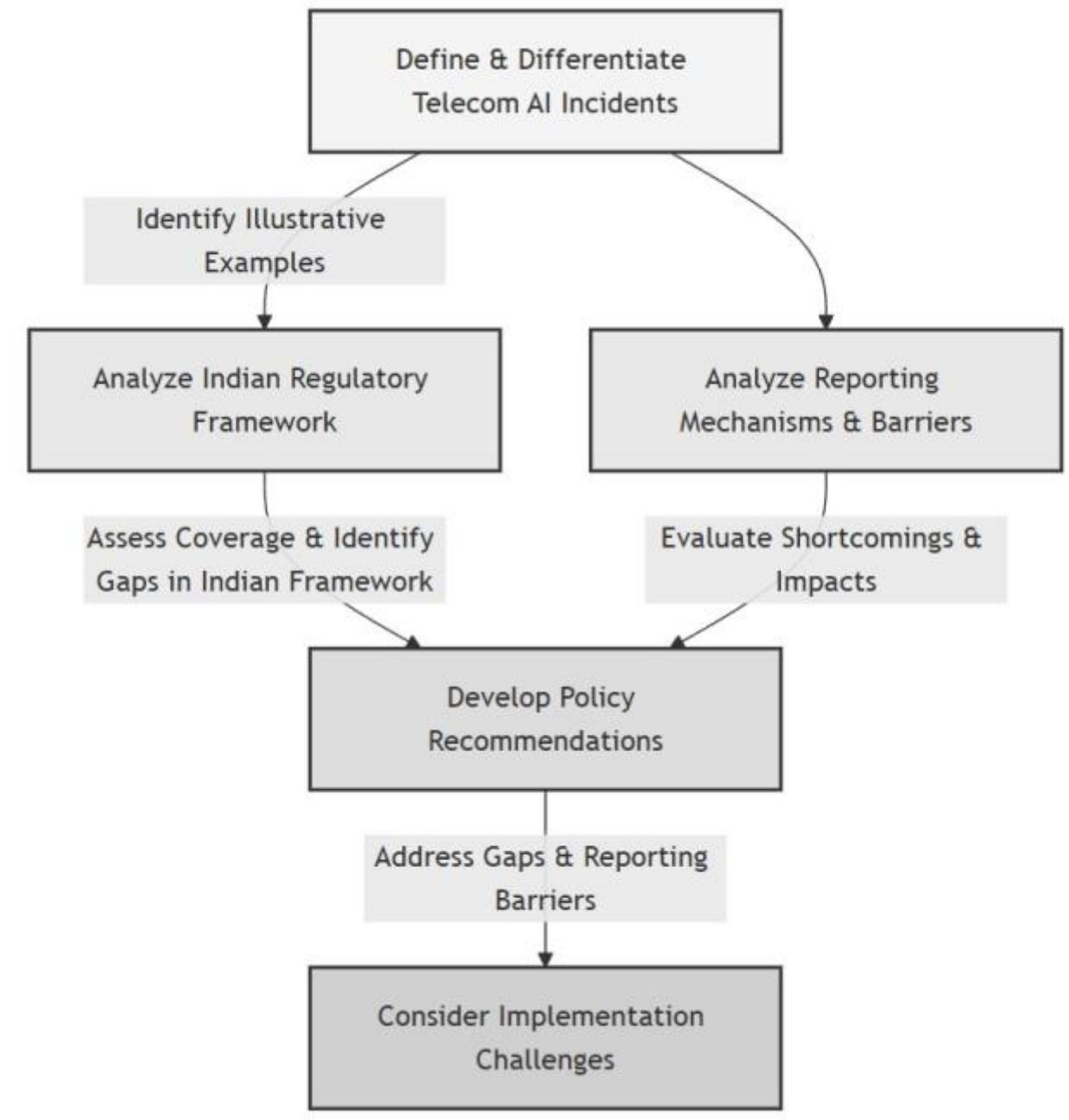

Figure 1: Research Methodology for AI Incident Reporting in Telecommunications

The need for such a specific definition is not merely theoretical. While telecom operators are often reluctant to publicly disclose such incidents due to commercial sensitivities and reputational risk, a growing body of industry analysis highlights this emerging challenge. The World Economic Forum warns that as AI is deployed across telecom functions, communication service providers must address risks of "bias, discrimination, liability" and "inadequate outcomes", such as in automated customer service, which fall outside traditional security paradigms [5]. For example, an AI-powered anti-spam system, such as that used by Bharti Airtel, could inadvertently block legitimate communications [5], creating a fairness and service quality issue rather than a security breach. Scholarly research has empirically demonstrated the existence of algorithmic bias in network optimization models [3], while other industry analysis points to risks from operational issues like model drift [4]. Furthermore, lessons from cybersecurity events like the T-Mobile incident reveal underlying operational vulnerabilities in AI systems, such as flawed decision-making in network control, that could also lead to non-malicious failures [37]. These examples underscore the existence of a class of AI induced failures focused on performance, fairness, and operational integrity that necessitate a distinct classification.

AI incidents in telecommunications encompass a range of failures, malfunctions, and unintended consequences resulting from the deployment or operation of AI systems within networks and services. This section proposes a domain-specific definition of telecommunications AI incidents, drawing from but extending beyond existing frameworks such as those in the OECD and the EU AI Act. It also presents illustrative examples and explores the relationship between such incidents and conventional cybersecurity threats and data protection violations, highlighting both areas of overlap and key differences.

### *3.1. Proposed Definition of a Telecommunications AI Incident*

The increasing deployment of AI systems within telecommunications networks necessitates a clear and specific definition of what constitutes an AI incident in this domain. Current definitions, such as those for cybersecurity incidents, do not adequately capture the range of potential risks posed by AI technologies, including unintended biases, vulnerabilities, data protection breaches, and unpredictable outcomes. To address this gap, the following definition is proposed:

*Telecommunications AI Incident means any event, circumstance, or malfunction involving the deployment or operation of AI systems within telecommunication networks or services that, whether through malicious intent or unintentional actions, directly or indirectly leads to: (a) disruption, degradation, or manipulation of telecommunication functions; (b) unauthorized access, alteration, or misuse of network resources or data; (c) the introduction of bias, vulnerabilities, or unpredictable outcomes in automated processes; or (d) harm to individuals, property, or the environment.*

The primary purpose of articulating such a comprehensive definition is to provide a clear legal basis for regulation. By formally incorporating this broader definition into relevant laws and rules, regulators can ensure that the full spectrum of AI incidents in telecommunications gets covered, including those operational and fairness related harms currently outside the scope of cybersecurity mandates. This closes the existing gap and leaves no ambiguity for telecom operators regarding their reporting obligations.

The proposed definition is derived from a critical analysis of existing definitions of AI incidents and cybersecurity incidents, emphasizing the unique risks posed by AI systems in the telecommunications sector. While cybersecurity incidents typically involve unauthorized access or data breaches, AI incidents encompass a broader spectrum of risks. These include not only security-related issues but also unintentional consequences arising from flaws in AI models, such as biases in automated decision making [3, 4], vulnerabilities due to lack of robustness, and unpredictable behaviors stemming from autonomous adaptation. This expanded scope is vital in telecommunications, where AI introduces tightly coupled dependencies between system components, alongside user-facing decisions that carry operational and ethical risks.

The definition draws upon frameworks such as the OECD's description of AI incidents, which include a wide array of harms. The OECD defines an "*AI incident*" as:

> *"an event, circumstance or series of events where the development, use or malfunction of one or more AI systems directly or indirectly leads to any of the following harms: (a) injury or harm to the health of a person or groups of people; (b) disruption of the management and operation of critical infrastructure; (c) violations of human rights or a breach of obligations under the applicable law intended to protect fundamental, labour and intellectual property rights; (d) harm to property, communities or the environment"* [38].

Similarly, the EU AI Act outlines serious incidents as those leading to significant harm or operational disruptions. As per the EU AI Act, a "*serious incident*" means:

> *"an incident or malfunctioning of an AI system that directly or indirectly leads to any of the following: (a) the death of a person, or serious harm to a person's health; (b) a serious and irreversible disruption of the management or operation of critical infrastructure; (c) the infringement of obligations under Union law intended to protect fundamental rights; (d) serious harm to property or the environment"* [39].

Admittedly, these high-level definitions are commendably broad and cover significant harms, including disruptions to critical infrastructure. However, their high reporting thresholds and focus on severe, tangible outcomes limit their applicability to the full spectrum of telecom-specific operational issues. The EU AI Act, for instance, requires a "serious and irreversible disruption" to critical infrastructure to trigger a mandatory report. This high bar would fail to capture many critical AI incidents in telecoms, such as systemic service degradation due to model drift, discriminatory quality of service arising from algorithmic bias, or inefficient network management causing widespread but non-critical performance issues. These operational incidents, while not "serious" in the catastrophic sense of the EU AI Act, are precisely the types of failures that erode trust and resilience over time and therefore must be systematically tracked. The OECD and EU definitions, while essential, are not a substitute for a domain-specific framework designed to capture this operational context.

Therefore, the proposed definition incorporates elements specific to the telecommunications domain, highlighting the dual nature of AI incidents: they can be either malicious or unintentional, and they may result in a range of adverse outcomes, including, but not limited to, cybersecurity threats and data protection violations. This distinction is essential to ensure comprehensive incident reporting and risk management practices in telecommunications policy, addressing the multifaceted nature of AI-induced risks.

*3.2. Illustrative Telecommunications AI Incidents*

The following examples illustrate the spectrum of AI incidents, particularly those that may not meet the threshold for, or fall within the scope of, traditional cybersecurity or data breach reporting obligations but still pose a significant risk to service quality, fairness, and user trust. The types of operational incidents described below, such as unfair resource allocation and performance degradation from model drift, are increasingly recognized in scholarly and industry analysis [3, 4]. AI-related incidents in telecommunications can arise in diverse forms depending on the nature of the system, use case, and regulatory context. These incidents may be grouped into three broad categories: (1) incidents that are neither cybersecurity nor data protection violations; (2) incidents that qualify as cybersecurity incidents due to malicious exploitation or vulnerabilities; and (3) incidents involving data protection breaches due to the mishandling of personal data.

A. *AI incidents not classified as cybersecurity or data protection violations*
   (a) Erroneous AI-based network optimization degrading service quality in specific regions.
   (b) AI-driven traffic management systems prioritizing low-value traffic, resulting in inefficient or unfair bandwidth allocation.
   (c) Predictive maintenance systems generating false alerts or overlooking real hardware faults, leading to service outages or increased operational costs.
   (d) Self-learning AI models adapting unpredictably to user demand patterns, causing instability or service denial in certain network segments.
   (e) AI-induced configuration errors affecting Quality of Service (QoS) without compromising security or personal data.
   (f) Profiling-based service differentiation using non-sensitive user data such as handset brand or operating system, potentially affecting service levels.
   (g) Biased recommendation systems affecting access to promotional telecom offers or customer support prioritization.
   (h) AI models scheduling software upgrades or network reboots at inappropriate times, disrupting services during peak usage periods.
   (i) Resource allocation models misinterpreting seasonal or regional demand trends, resulting in under-provisioning in high-traffic areas.
   (j) Faulty AI-driven prioritization of network restoration tasks during outages, delaying recovery for critical or high-need users.
   (k) AI-driven language translation or voice recognition errors in customer service interactions, leading to miscommunication or user frustration without a security or data breach.
   (l) Overfitting in AI models trained on limited datasets, triggering overreactive network scaling decisions and increasing operational costs.

B. *AI-Driven Cybersecurity Incidents*
   (a) Exploitation of AI-based intrusion detection or traffic monitoring systems using adversarial inputs to bypass detection.
   (b) Misclassification by AI-controlled firewalls, resulting in unauthorized access or blocking of legitimate traffic.
   (c) Use of generative AI to generate highly convincing phishing content targeting telecom staff or customers.
   (d) Manipulation of AI-powered anomaly detection tools to obscure lateral movement or data exfiltration during an attack.
   (e) Deployment of compromised AI agents in edge or cloud systems, introducing backdoors or enabling remote code execution.

C. *AI-Related Data Protection Incidents*
   (a) Inferring sensitive user attributes, such as location, habits, or preferences, without a lawful basis or explicit user consent.
   (b) Enabling re-identification of individuals due to training AI models on improperly anonymized datasets.
   (c) Violating data subject rights due to automatic decision-making systems lacking transparency or mechanisms for user contestation.
   (d) Disclosing private information through AI generated outputs in chatbots, customer service logs, or automated messaging systems.
   (e) Processing personal communications metadata by AI systems beyond their stated purpose, without sufficient legal safeguards.

*3.3. Distinction between AI incidents, cyberattacks, and data breaches*

The illustrative AI incidents in telecommunications discussed in the previous subsection demonstrate that not all AI incidents qualify as cyberattacks or data protection violations. Many AI incidents stem from flawed decision-making, performance degradation, bias, lack of explainability, or unsafe behavior, none of which necessarily involve malicious intent or unlawful data processing. These incidents can arise from limitations in AI design, incomplete training data, or misaligned deployment goals, rather than from adversarial attacks or data misuse.

Furthermore, while some of these issues, such as bias or performance degradation, can exist in non-AI operational systems, AI introduces qualitatively different challenges that traditional operational frameworks, such as Service Level Agreements (SLAs), are not designed to address. The key differentiators are scale, opacity, and autonomy. AI operates at a scale and velocity that can amplify a minor flaw into a massive, cascading failure far faster than a human-monitored system [40]. Critically, the "black box" nature of many complex models makes root cause analysis exceptionally difficult; unlike a traditional software bug, it may be impossible to determine precisely why an AI made a specific biased decision. Finally, the ability of AI systems to learn and adapt autonomously can lead to unpredictable emergent behaviors or model drift, creating risks that were not present in the system's original design and are not captured by static operational assessments.

At the same time, some AI incidents may indeed manifest as cybersecurity incidents, such as when an attacker exploits a vulnerability in an AI-based intrusion detection system, or when generative AI is misused to produce sophisticated phishing campaigns targeting telecom networks. Similarly, AI-related data protection violations can occur, such as when AI models process personal data without an appropriate legal basis, infer sensitive user information through profiling, or fail to uphold data subject rights due to automation. However, not all cyberattacks or data breaches necessarily involve AI. Many cyber incidents in telecom arise from conventional vectors like malware, misconfigured firewalls, or social engineering, which do not involve any AI system. Likewise, numerous data breaches result from human error, insider threats, or unsecured databases, events that may unfold independently of any AI component in the network.

There exists a distinct subset of AI incidents in telecommunications that are neither cybersecurity incidents nor data protection breaches. These incidents do not involve malicious attacks or unlawful handling of personal data but nonetheless arise from flawed, unpredictable, or biased behavior of AI systems deployed in network operations. Such cases may lead to degraded service quality, increased operational costs, or erosion of user trust, yet fall outside the traditional scopes of cybersecurity or data privacy regulation. Recognizing and separately tracking this broader category of AI incidents is essential for developing comprehensive governance mechanisms that reflect the full spectrum of AI-related risks in telecom infrastructure.

Figure 2 illustrates the overlap between AI incidents, cyberattacks, and data breaches.

## 4. Analysis of India's Regulatory Framework and its Gaps

Having established a conceptual definition for *Telecommunications AI Incidents* in Section 3, this section analyzes India's existing legal landscape to assess its adequacy in addressing these novel risks. In the absence of a horizontal, overarching AI law, the analysis focuses on three key pillars: the foundational cybersecurity rules, the modern telecommunications framework, and the new data protection regime, to determine if their scope extends beyond traditional security threats to cover the full spectrum of AI-induced failures.

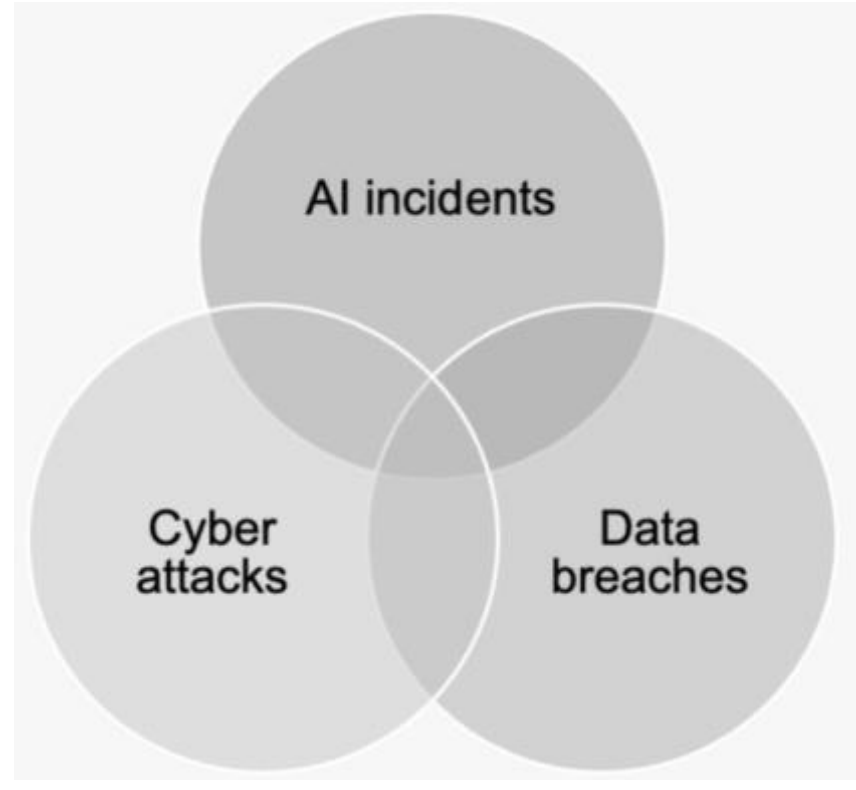

Figure 2: AI incidents, cyberattacks, and data breaches

### *4.1. The Foundational Framework: The IT Act and CERT-In Rules*

India's baseline for cybersecurity incident management is established by *The Information Technology Act, 2000* (as amended in 2008) [41]. Under this Act, the *Information Technology (The Indian Computer Emergency Response Team and Manner of Performing Functions and Duties) Rules, 2013* (CERT-In Rules) create the mandate for reporting [42].

The rules provide two key definitions that frame the reporting obligation. Rule 2(g) broadly defines a "*cyber incident*" as:

> *"any real or suspected adverse event that is likely to cause or causes an offence or contravention, harm to critical functions and services across the public and private sectors by impairing the confidentiality, integrity, or availability of electronic information, systems, services or networks resulting in unauthorised access, denial of service or disruption, unauthorised use of a computer resource, changes to data or information without authorisation; or threatens public safety, undermines public confidence, have a negative effect on the national economy, or diminishes the security posture of the nation;"* [42].

Rule 2(h) provides a more specific definition for a "*cyber security incident*" as:

*“any real or suspected adverse event in relation to cyber security that violates an explicitly or implicitly applicable security policy resulting in unauthorized access, denial of service or disruption, unauthorised use of a computer resource for processing or storage of information or changes to data, information without authorisation”* [42].

Admittedly, the language of Rule 2(g) is broad enough to cover a catastrophic AI-induced failure, such as a complete network outage, which would clearly constitute an ”impairment of availability.” However, the framework’s entire conceptual model is rooted in security paradigms. This is evidenced by the Annexure to the rules, which lists mandatory reportable incidents such as *“Targeted scanning/probing of critical networks/systems,” “Compromise of critical systems/information,”* and *“Unauthorised access of IT systems/data.”*

This security-centric focus creates a significant gap. The legal triggers for reporting are tied to concepts like “unauthorized access,” “denial of service,” or violations of a “security policy.” This framework is not designed to address AI incidents that do not cause a full ”disruption” but still result in significant harm. For example, an AI model exhibiting algorithmic bias that provides consistently degraded, but still functional, service to a specific demographic would not be a reportable incident under these rules. Similarly, gradual performance degradation due to model drift, which erodes service quality over time without causing a discrete outage, falls outside this legal framework. These are precisely the types of AI-specific operational harms that the CERT-In rules are not designed to capture.

### *4.2. The Modern Framework: The Telecommunications Act, 2023 and its Rules*

The recently enacted *Telecommunications Act, 2023* represents a significant modernization of India’s telecom governance [15]. The Act grants the Central Government broad powers to establish measures for ensuring “cyber security for telecommunication services and telecommunication networks” (Sections 19(e) and 22). However, the Act itself does not define what constitutes a reportable incident, nor does it explicitly mention AI systems or address how algorithmic failures unrelated to security could affect network services.

The implementing rules issued under the Act reveal a consistent and exclusive focus on security. First, the *Telecommunications (Telecom Cyber Security) Rules, 2024* mandate strict incident reporting timelines, requiring entities to report incidents within six hours [43]. The scope of this obligation is defined by Rule 2(e), which states that a *“security incident”* means *“an event having real or potential risk on telecom cyber security.”* This definition tethers the reporting requirement directly and exclusively to cybersecurity risks.

Second, this security-centric approach is mirrored in the *Telecommunications (Critical Telecommunication Infrastructure) Rules, 2024* [44]. These rules also mandate a six-hour reporting timeline for “security incident(s)” affecting designated critical infrastructure (Rule 7(1)(j)). While critically important for protecting the nation’s core networks, these rules again do not expand the definition of a reportable incident beyond the security domain.

This narrow and consistent focus across the entire modern telecommunications framework creates a significant regulatory blind spot. The legal obligation to report is triggered only by events that pose a “risk on telecom cyber security.” This leaves no mechanism to address the full spectrum of telecommunications AI incidents, such as those rooted in algorithmic bias, model drift, or flawed automation, that can degrade service quality and erode user trust without posing a direct security risk.

### *4.3. The Data Protection Lens: The DPDP Act, 2023*

The third pillar of India’s digital governance is the *Digital Personal Data Protection Act, 2023* (DPDP Act) [16]. The rules to operationalize this Act are still being finalized; the draft *Digital Personal Data Protection Rules, 2025*, for instance, were recently released for public consultation to elaborate on the specifics of security safeguards and breach notifications [45]. This landmark legislation establishes a consent-based framework for processing personal data and mandates the reporting of any “personal data breach.” However, its jurisdiction is triggered only when the confidentiality, integrity, or availability of *personal data* is compromised.

This creates another clear boundary. A telecommunications AI incident that impacts network performance or operational integrity without involving a personal data breach falls entirely outside the purview of the DPDP Act. For instance, an AI-driven traffic management system causing network congestion, or a biased algorithm providing slower speeds to a certain demographic, would not constitute a reportable personal data breach and would therefore be unaddressed by this law.

### *4.4. Identified Regulatory Gaps in the Indian Framework*

This detailed analysis of India’s legal landscape reveals four critical gaps that hinder effective AI incident management in the telecommunications sector:

1. *Limited Scope of Telecom Laws*: The *Telecommunications Act, 2023* and its associated rules continue to center on cybersecurity, with limited recognition of AI-specific operational risks.
2. *Narrow Coverage of Broader Laws*: The CERTIn rules and the DPDP Act address AI risks only when they result in a security violation or a personal data breach, respectively. Incidents of performance degradation, biased decision-making, or service manipulation remain outside the purview of these mandates.

3. *A Misalignment with New Technological Realities*: The existing legal instruments, while robust for their intended purpose, were not designed to address the unique failure modes of complex, autonomous AI systems.
4. *Focus on Incident Response, Not Systemic Learning*: Where incident reporting exists, it is reactive and centered on mitigating immediate security threats or data breaches, with insufficient emphasis on systemic learning from a broader class of operational AI failures.

In sum, while India has built a strong framework for managing cybersecurity incidents and personal data breaches, the integration of AI introduces fundamentally different challenges that existing regulations do not adequately address. Recognizing and closing these gaps is essential for ensuring that AI-induced failures in critical telecom systems are documented, understood, and mitigated.

## 5. Barriers and Limitations in AI Incident Reporting

Effective AI incident reporting is essential for identifying risks, improving accountability, and strengthening resilience in critical sectors. However, various barriers hinder organizations from reporting AI incidents, including regulatory uncertainties, liability concerns, and operational challenges. Additionally, existing AI incident repositories, while useful, are not designed to capture sector-specific failures, particularly in telecommunications and critical infrastructure. This section examines the key barriers to AI incident reporting and analyzes the limitations of current repositories in addressing telecom specific AI risks.

### *5.1. Barriers to AI incident reporting*

While critical for governance and risk management, several barriers hinder AI incident reporting. Prior studies have examined barriers in the context of cybersecurity incident reporting [46, 47, 48, 49], with some research addressing similar challenges in medical incident reporting [50, 51, 52], but AI-specific barriers remain underexplored.

While many of these barriers are common to any incident reporting regime, several are significantly exacerbated or take on a unique character in the context of AI. For instance, the "black box" nature of many complex AI models makes the *Technical Challenge* of root cause analysis exceptionally difficult compared to conventional software bugs. Similarly, given the intense public scrutiny and hype surrounding AI, the *Trust and Reputation Concern* over disclosing an "AI failure" is often heightened. The following subsection extends insights from other fields to AI incident reporting, contextualizing these challenges for the telecom sector.

A. Operational and Resource Constraints
   - (a) Limited financial and human resources, especially for smaller organizations.
   - (b) Capacity constraints in processing and responding to incidents.
   - (c) Time constraints and competing priorities, leading to a sense of futility in reporting.
   - (d) Lack of trained personnel with expertise in AI incident assessment and response.
   - (e) Lack of organizational support and leadership commitment to promoting a reporting culture.
   - (f) Internal resistance to change and low motivation among personnel to actively participate in incident reporting.

B. Technical Challenges
   - (a) Lack of standardization and interoperability in incident reporting frameworks.
   - (b) Vague or inconsistent terminology causing confusion and misinterpretation.
   - (c) Difficulty in determining accuracy, validating quality, and ensuring timely reporting.
   - (d) Balancing automation with contextual understanding in incident classification.
   - (e) Lack of a centralized system for secure information exchange and communication.
   - (f) Managing redundant or irrelevant data while maintaining reporting efficiency.

C. Trust and Reputation Concerns
   - (a) Fear of reputation damage or loss of client confidence due to disclosed incidents.
   - (b) Reports being perceived as indicators of incompetence, discouraging participation.
   - (c) Free riding, where some entities benefit from shared intelligence without contributing.
   - (d) Lack of trust among stakeholders, limiting open and transparent reporting.
   - (e) Fear of negative consequences from co-workers or supervisors discouraging reporting.

D. Legal and Policy Barriers
   - (a) Uncertainty regarding privacy, liability, and regulatory compliance.
   - (b) Divergent legal frameworks across jurisdictions complicating adherence.
   - (c) The risk of violating privacy or antitrust laws when sharing incident data.
   - (d) Fear of regulatory reprisals or legal consequences for disclosing incidents.
   - (e) Government over-classification restricting transparency and collaboration.
   - (f) Concerns over unintentionally exposing sensitive data.

E. Accessibility and Awareness Issues

(a) Inaccessibility of reporting platforms and lack of standardized submission channels.

(b) Limited awareness on AI incident reporting.

(c) Lack of structured training programs for capacity building in incident reporting.

(d) Absence of incentives or mandates to encourage proactive reporting.

These challenges highlight the need for AI-specific adaptations to incident reporting frameworks, ensuring resilience, accountability, and trustworthiness in AI driven systems.

### *5.2. Gaps in AI Incident Reporting Mechanisms for Telecommunications*

Existing AI incident repositories, such as AIID and AIAAIC, document a wide range of AI-related failures, harms, and controversies. While these repositories provide valuable insights, they primarily focus on consumer-facing applications and social impacts, with limited representation of incidents in critical infrastructure, including telecommunications. To assess their relevance to telecom-specific AI incident reporting, this work reviewed several repositories and identified key gaps in their scope, coverage, and methodology.

Several repositories have been developed to document AI-related incidents, each with distinct scopes and methodologies, but their general-purpose nature limits their applicability to the telecom sector. The AI, Algorithmic, and Automation Incidents and Controversies (AIAAIC) Repository compiles incidents broadly related to AI, algorithms, and automation, emphasizing ethical, social, and technical impacts [35]. The AI Incident Database (AIID) is inspired by aviation and cybersecurity incident databases and records real-world harms or near harms caused by AI systems [34]. Both these repositories rely on open submissions and consequently focus on publicly visible, often consumer-facing, harms. An analysis of their contents shows that a large proportion of reported incidents involve social media platforms or demographic bias, rather than the technical, operational failures within critical infrastructure [21]. The AI Incidents Monitor (AIM), developed by the OECD.AI expert group, uses machine learning to identify and classify AI related incidents by scraping public news sources [53], a method which, by definition, cannot capture the internal and often commercially sensitive operational incidents that are a key concern for telecom resilience. Other initiatives like the AI Vulnerability Database (AVID) also contribute to broader accountability [54]. While these databases are invaluable for tracking broader AI accountability, they lack the specific technical taxonomies and sectoral focus required to systematically capture and analyze AI failures in telecommunications.

Despite the growing number of AI incident databases, significant gaps remain in systematically capturing and analyzing failures in telecommunications. An examination of reported incidents and prior research reveals the following seven key limitations:

1. *Limited sector-specific coverage:* Most AI incident databases focus on general AI failures, overlooking telecom-specific issues such as AI-driven network disruptions, misconfigurations, and predictive maintenance failures. Critical digital infrastructure sectors, including telecommunications, remain underrepresented [21].
2. *Voluntary reporting and underreporting:* These databases operate as private efforts without regulatory mandates, leading to significant underreporting since organizations are not obligated to disclose incidents. The absence of oversight and reliance on voluntary contributions also result in a lack of accountability, further limiting transparency and comprehensive incident tracking [7, 25].
3. *Inconsistent definitions and lack of standardized taxonomies:* The absence of universally accepted definitions and taxonomies for AI incidents makes categorization and comparative analysis challenging. Some reported cases in these repositories do not align with widely accepted AI incident definitions [27].
4. *Lack of granularity:* Existing repositories fail to capture critical details such as root causes, severity levels, and AI system versions, limiting their usefulness for in-depth analysis and effective mitigation [27, 36].
5. *Reporting bias and arbitrariness:* Reports are disproportionately sourced from a small group of individuals and a few Western newspapers, leading to arbitrariness in the incident selection, assessment, and classification, as well as potential biases and inconsistencies [25, 6].
6. *Fragmentation and lack of interoperability:* Different repositories use distinct structures and schemas, making it difficult to aggregate, compare, and analyze incidents across platforms [36].
7. *Geographical imbalance:* AI incident data is heavily skewed toward North America and Europe, with minimal representation from developing and underdeveloped regions. This creates a significant blind spot regarding the specific challenges and incident types that may arise in rapidly growing markets like India, hindering a global understanding of AI risks [6].

AI incident reporting faces both systemic barriers and gaps in existing repositories. Organizational hesitancy, regulatory inconsistencies, and lack of incentives hinder transparent reporting. Meanwhile, existing repositories remain general-purpose, with limited sectoral focus, inconsistent taxonomies, and voluntary participation, leading to underreporting in critical infrastructure. Overcoming these challenges in the Indian context requires a structured, sector-specific framework with regulatory backing to enhance AI risk monitoring in telecommunications.

## 6. Policy Recommendations for AI Incident Reporting in Telecommunications

The analysis in Section 4 demonstrated that while India has robust frameworks for cybersecurity and data protection, they contain significant gaps regarding the novel risks posed by AI. In the absence of a comprehensive, horizontal AI law in India, and with no clear roadmap to introduce such a law in the near future, strengthening existing sectoral regulations is the most pragmatic and effective path forward. This section outlines eight targeted policy recommendations designed to bridge the identified regulatory gaps by integrating AI incident management directly into India's telecommunications governance framework.

### *6.1. Integrate AI Incident Reporting into Telecom Regulations*

To enable systematic data collection of AI incidents, the scope of the Rules being notified under the *Telecommunications Act, 2023* should be expanded to mandate AI incident reporting for high-risk incidents. This expansion must begin with the formal adoption of a broader definition of a reportable "incident," such as the one proposed in this paper, so that the full spectrum of AI-specific risks is explicitly brought under regulatory oversight. This would require telecom service providers to report such incidents as defined by new regulatory risk classifications. Additionally, regulations should facilitate voluntary reporting by affected customers and individuals to ensure a more comprehensive dataset. While cybersecurity incident reporting primarily aims to prevent malicious attacks, AI incident reporting should focus on gathering structured data to improve AI models, enhance network resilience, and prevent repeat failures.

The *Telecommunications Act, 2023* already provides an avenue for such innovation. Its provision for Regulatory Sandboxes [15] could be leveraged to pilot AI incident reporting frameworks, enabling controlled data sharing for research and regulatory learning, and ensuring that reporting obligations contribute to both security and technological advancements in telecommunications.

### *6.2. Designate a Nodal Agency for AI Incident Management*

The central government should designate a sector-specific nodal agency for AI incident management to ensure structured oversight and enforcement. To avoid governance fragmentation, this should not be a new body, but rather a function designated within an existing organization with relevant technical expertise and regulatory authority, such as the Telecommunication Engineering Centre (TEC) or the Telecom Regulatory Authority of India (TRAI). This nodal agency would be responsible for maintaining a comprehensive telecommunication AI incident repository, issuing guidelines, and ensuring compliance with reporting protocols.

The responsibilities of the designated nodal agency would include, but are not limited to:

1. Issuing guidelines and standards for structured reporting and safe disclosure of telecommunication AI incidents while safeguarding personal and proprietary data.
2. Developing a comprehensive framework for AI incident reporting, ensuring consistency with global best practices.
3. Establishing and maintaining a secure, anonymized online telecommunication AI incident repository for regulatory oversight and research.
4. Collaborating with telecom service providers, vendors, academia, and research organizations to analyze incidents, identify recurring risks, and develop mitigation strategies.
5. Imposing penalties on telecom entities that fail to comply with AI incident reporting and management protocols.
6. Undertaking additional responsibilities as notified by the government over time.

The nature of this nodal authority is crucial to its effectiveness. A government department may lack the necessary technical expertise and could face conflicts of interest, while a private entity may not be perceived as neutral or authoritative enough to manage sensitive data. Therefore, the most suitable approach would be to designate a specialized technical organization operating under some government oversight. This structure would allow the authority to engage skilled expertise, maintain trust among stakeholders, and enforce compliance while ensuring confidentiality in AI incident management.

By institutionalizing AI incident monitoring through a dedicated authority, telecom governance structures can enhance sector-wide resilience, issue early warnings, and develop evidence-based policy recommendations to mitigate AI-related risks in telecommunications.

### *6.3. Mandate Risk Assessment for AI Systems in Telecom*

Regulations should establish distinct and measurable AI risk levels, such as low, limited, high, and unacceptable, to ensure proportionate oversight. The EU AI Act follows a similar approach, providing a useful international reference for categorizing AI risks [55]. This approach also aligns with the recommendations of the Telecom Regulatory Authority of India (TRAI), which, in its July 2023 paper on AI, proposed a risk-based regulatory framework for the sector [56]. Building on this, telecom service providers should be required to conduct periodic risk assessments and risk grading for all AI applications deployed in telecom networks. These assessments must identify system vulnerabilities, potential failures, and unintended consequences to determine their risk classification. Risk evaluations should be conducted at predefined intervals, before major system upgrades, and as part of regulatory audits to ensure compliance and continuous risk monitoring.

### *6.4. Introduce Incentives for AI Incident Reporting*

Regulations should combine mandatory reporting for high-risk incidents with incentives for reporting compliance to strengthen AI incident reporting. While telecom service providers must report high-risk incidents as per regulatory mandates, structured incentives can encourage timely and comprehensive disclosures. Additionally, incentives should promote voluntary reporting by telecom equipment manufacturers, vendors, consumers, and other stakeholders.

Regulatory incentives may include liability protections for good-faith reporting, compliance benefits for proactive risk management, and access to anonymized industry-wide insights for improved risk mitigation. Establishing clear incentive mechanisms will promote transparency, enhance incident data collection, and strengthen the resilience of AI-driven telecom systems.

### *6.5. Ensure Transparency and Data Protection in AI Incident Reporting*

AI incident reporting must protect sensitive information while enabling meaningful oversight. Anonymization protocols, in line with the principles of the *DPDP Act, 2023*, should prevent the exposure of personally identifiable information, proprietary business data, and the identity of reporting entities. Structured disclosure formats should be established to balance transparency with confidentiality, ensuring that reported incidents remain useful for analysis while mitigating risks of misuse. Secure reporting channels and controlled data-sharing mechanisms should facilitate cross-stakeholder collaboration while maintaining strict privacy safeguards. Protecting the identities of individuals or organizations submitting reports will encourage disclosures without fear of retaliation or reputational harm, strengthening the effectiveness of AI incident reporting frameworks.

### *6.6. Standardize Processes and Taxonomy*

The lack of standardized AI incident classification and reporting formats limits effective data aggregation and analysis. A unified incident reporting schema should be established to ensure consistency across AI incident databases, enabling seamless data integration and cross-sectoral comparisons [36]. A primary responsibility of the designated nodal agency should be to develop a standardized taxonomy, specifically tailored for AI incidents in telecommunications. This will support meaningful trend analysis and risk mitigation efforts by capturing key attributes such as root causes, severity levels, and affected network functions. Aligning these processes with global AI governance initiatives will enhance interoperability and improve trust in AI-driven telecom systems.

### *6.7. Integrate AI Risk Assessments into Telecom Equipment Certification*

Telecom equipment in India undergoes rigorous conformity assessment under the Mandatory Testing and Certification of Telecom Equipment (MTCTE) regime. This framework, however, primarily focuses on parameters like Electromagnetic Interference (EMI), Electromagnetic Compatibility (EMC), safety, and traditional telecom security [57]. As AI-driven functionalities become integral to telecom infrastructure, the MTCTE framework should be expanded to include AI-specific assessment criteria.

AI-related parameters such as fairness, robustness, and security risks should be incorporated into existing conformity assessments to ensure that AI-powered network components meet predefined reliability and safety standards before deployment. Aligning these certifications with international AI risk evaluation standards, including emerging frameworks for fairness assessment [58], will enhance consistency, interoperability, and trust in AI-enabled telecom systems.

### *6.8. Strengthen Global Cooperation on AI Incident Reporting*

AI incidents in telecommunications often have cross-border implications, requiring structured international cooperation. The ITU already supports the establishment of Computer Incident Response Teams (CIRTs) for cybersecurity as per ITU-D Resolution 69 and ITUT Resolutions 50 and 58 [29, 30, 31]. This mandate could be extended to cover AI incidents beyond cybersecurity risks. By developing a robust national framework for AI incident reporting, India can position itself as a key contributor to these global standard-setting discussions. Global coordination through such UN-backed bodies can facilitate standardized AI incident reporting and improve response strategies. Strengthening collaboration between telecom regulators, cybersecurity agencies, and AI governance institutions will enhance incident tracking and mitigation on a global scale.

Implementing these recommendations, as summarized in Table 1, will enhance the detection, analysis, and prevention of AI-related failures in telecommunications. A robust and trustworthy telecom ecosystem requires a regulatory framework that mandates reporting, incentivizes compliance, and encourages international cooperation.

## 7. Discussion and Future Directions

This study examined the need for AI incident reporting in telecommunications policy to enhance resilience. The findings, derived from a focused case study of India, address the four research questions and contribute to defining telecommunications AI incidents, identifying specific regulatory gaps in the Indian context, and proposing actionable policy measures.

The first research question asked how AI incidents in critical telecommunications infrastructure relate to traditional cybersecurity and data protection incidents, and why their reporting is crucial for resilience. This study provided a clear definition of *telecommunications AI incidents*, establishing them as a distinct category of risk that extends beyond, while partially overlapping with, general AI failures as well as cybersecurity breaches and data protection violations. While cybersecurity incidents typically involve external threats such as unauthorized access or malware, and data protection incidents concern the misuse of personal data, telecommunications AI incidents also include risks arising from the internal behavior of AI systems. These include algorithmic misjudgments, unintended biases, or unpredictable outputs, which may not constitute a security or privacy breach. Examples include AI-driven traffic mismanagement, errors in fault prediction, or unfair outcomes in automated provisioning decisions. By articulating a clear definition, this study enables policymakers and industry stakeholders in India and beyond to recognize these distinct risks, forming the basis for targeted regulatory interventions. Without such a definition, AI-related failures risk being misclassified, limiting the ability to develop effective mitigation strategies.

The second research question examined whether India's current telecom regulations adequately address AI related risks. This study identified four key gaps in the existing legal and regulatory framework. It found that foundational laws like the *IT Act* and modern frameworks like the *Telecommunications Act, 2023*, continue to focus narrowly on cybersecurity. Broader legal instruments, such as the *DPDP Act, 2023*, likewise cover AI-related failures only if they directly result in a personal data breach. Furthermore, these frameworks were not designed for the unique failure modes of AI, limiting their relevance to the evolving technological landscape. Even where incident reporting obligations exist, the emphasis remains largely on immediate response and mitigation, with insufficient focus on systematically collecting data for systemic learning. These findings highlight a critical regulatory lag: AI incidents that do not cause direct security or privacy harm may still undermine service quality and public trust, yet they often remain underreported. This study, therefore, underscores the need for Indian regulations to evolve, not merely by expanding scope, but by integrating AI risks as a distinct policy domain.

The third research question focused on identifying gaps, barriers, and challenges in AI incident reporting. This study identified two primary challenges in the Indian context: (1) significant barriers to incident reporting, such as the voluntary nature of disclosure and deepseated concerns over reputational harm in a competitive market; and (2) the absence of a structured, mandated AI incident reporting mechanism specific to the telecom sector. Unlike cybersecurity breaches, which require mandatory disclosure under CERT-In rules, AI-related operational failures remain largely undocumented. The analysis of existing global AI incident databases, such as AIID and AIAAIC, further revealed their inadequacy, citing a lack of telecom-specific focus and a geographical imbalance that underrepresents challenges in markets like India. This highlights that without a structured and mandatory national reporting framework, AI-driven failures will remain largely invisible to Indian regulators and researchers, impeding the development of robust resilience strategies.

Table 1: Policy recommendations for AI incident reporting in the Indian telecommunications context

| **Policy categories** | **Policy Recommendations** |
|---|---|
| Regulatory measures | Integrate AI incident reporting into telecom regulations |
| | Designate a Nodal Agency for AI Incident Management |
| Risk management and standardization | Mandate risk assessment for AI systems in telecom |
| | Standardize processes and taxonomy |
| | Integrate AI risk assessments into telecom equipment certification |
| Incentives and international cooperation | Provide incentives for AI incident reporting |
| | Enhance cross-border cooperation on AI incident reporting |
| | Ensure transparency and anonymization in AI incident data sharing |

The fourth research question examined regulatory and policy interventions that could enhance AI incident reporting. This study proposed actionable recommendations, the foremost being that rules under the *Telecommunications Act, 2023* should be updated to include AI incident reporting for high-risk cases. The paper advocates for a sector-specific nodal agency, housed within an existing government body like TEC or TRAI, to be responsible for collecting and managing AI incident data while ensuring confidentiality and industry cooperation. Additionally, it recommends establishing a standardized reporting framework and mandating regular risk assessments of AI systems. These, along with the other recommendations proposed in this paper, would improve transparency, facilitate early detection of AI-related risks, and strengthen the resilience of India's critical telecommunications networks.

### *7.1. Broader Implications and Replicability*

While this paper's analysis focuses on the Indian context, its findings have significant implications for a wide range of countries. India's situation, a rapidly modernizing telecom sector, new but not-yet-fully-implemented digital laws, and the absence of a horizontal AI Act, is not unique. It mirrors the reality in numerous nations across Asia, Africa, and Latin America, with various non-EU jurisdictions like Australia, Nigeria, and Singapore grappling with the challenge of adapting existing sectoral laws to govern AI risks in the absence of a comprehensive, overarching AI Act. Therefore, the policy recommendations proposed, while tailored to India's specific legal instruments, are designed as a replicable blueprint. The core principle of embedding AI incident reporting into existing sectoral regulations provides a pragmatic and immediate strategy that can be adapted by other regulators globally who face this common challenge. While a detailed comparative analysis of the specific legal frameworks in these jurisdictions is beyond the scope of this paper, it remains a valuable and necessary direction for future research.

### *7.2. Limitations and Future Work*

While this study provides a policy-oriented foundation for AI incident reporting, several challenges, especially related to practical implementation, remain.

First, the absence of mandatory reporting frameworks means that empirical data on AI failures in the Indian telecom sector remains scarce. Without real-world reporting mechanisms, assessing the full impact of AI incidents is difficult. A valuable and pragmatic first step would be to conduct a pilot implementation of the proposed reporting framework, perhaps leveraging the "Regulatory Sandbox" provision of the Telecommunications Act, 2023, to test its effectiveness and operational feasibility before a full rollout.

Second, industry resistance to AI incident disclosure is a significant barrier. Telecom operators in competitive markets may be unwilling to share information about AI failures unless provided with legal protections or incentives. Future work should explore models of incentivized reporting suitable for the Indian context, such as confidential reporting to a trusted nodal agency or legal protections for good-faith disclosures. This would allow for outcome-based oversight focused on performance and fairness metrics, rather than requiring the disclosure of sensitive model architectures, thus balancing regulatory needs with commercial realities.

Third, as AI in telecommunications evolves globally, new types of incidents will inevitably emerge, such as failures in AI-powered beamforming, dynamic spectrum allocation, or network slicing. This study provides a policy framework for AI incident reporting as it stands today, but ongoing updates will be necessary. Future research worldwide will need to explore adaptive regulatory mechanisms that can accommodate these emerging technological risks, the resource constraints of different operators through phased or proportionate implementation schedules.

Finally, while this study focuses on AI incidents within telecommunications networks, a critical area for future research is the broader impact of telecom-related AI failures on interconnected sectors. Future work should examine these potential cascading effects across vital areas like autonomous transportation, industrial automation, and emergency communication systems. Given the increasing reliance of all AI applications on telecom infrastructure, understanding and mitigating these cross-sectoral risks will be crucial for strengthening national and global resilience.

## 8. Conclusion

The increasing integration of AI in telecommunications introduces novel risks that extend beyond traditional cybersecurity threats, potentially affecting network resilience and operational stability. While existing regulations emphasize protection against cyber incidents and data breaches, they lack explicit provisions for AI-specific failures such as algorithmic bias, unpredictable behavior, or systemic errors that may not constitute security breaches but still degrade service quality, introduce vulnerabilities, or erode user trust. The absence of a structured reporting mechanism for such incidents means that these failures, when they occur, remain undocumented, preventing the industry from learning from past occurrences and implementing preventive measures.. Systematically compiling AI incidents is crucial for identifying emerging patterns, improving AI system design, and strengthening telecom resilience.

This study contributes to enhancing telecommunications policy by addressing these challenges through a focused case study of India, a nation representative of many others with a rapidly digitizing economy but without a horizontal AI law. First, the paper establishes crucial conceptual clarity by defining *telecommunications AI incidents* and presenting a detailed typology of such events, which clarifies their relationship with general AI failures and expands their scope beyond traditional cybersecurity and data protection breaches, establishing AI risks as a distinct regulatory concern.

Second, using this lens, the paper identifies four major regulatory gaps in India's legal framework. It reveals that the current laws from the CERT-In Rules to the *Telecommunications Act, 2023*, and the *DPDP Act* are designed to respond to security policy violations and personal data breaches. This narrow focus leaves critical AI-specific operational issues unaddressed. For example, the frameworks do not cover incidents such as algorithmic bias leading to discriminatory service quality for certain user groups, or gradual performance degradation due to model drift, both of which erode user trust and operational efficiency without triggering a security alert. The study also highlights that these frameworks emphasize reactive incident response over proactive, systemic data collection.

Third, based on this analysis, the study proposes eight targeted and actionable policy measures. These recommendations provide a pragmatic roadmap for Indian policymakers, centered on integrating AI incident reporting into the existing telecom regulatory framework. Key proposals include mandating reporting for high-risk AI incidents; designating an existing government body as a nodal agency for telecom AI incident data collection and management; incentivizing voluntary reporting; and developing standardized reporting frameworks to ensure consistency and reliability in documenting AI incidents in telecommunications.

By addressing these aspects, this work lays the foundation for integrating AI incident reporting into telecommunications governance. Crucially, while the analysis is grounded in the Indian context, the findings and the proposed sectoral approach serve as a replicable blueprint for the many other countries in a similar position: governed by domain-specific laws and seeking to manage the risks of AI without a comprehensive, overarching legal framework. Ensuring that the full spectrum of AI related failures is systematically recorded and analyzed provides a vital and immediate strategy for enhancing the resilience, accountability, and trustworthiness of the telecom networks that underpin our global digital society.

## References

[1] X. Wang, X. Li, V. C. Leung, Artificial intelligence-based techniques for emerging heterogeneous network: State of the arts, opportunities, and challenges, IEEE Access 3 (2015) 1379–1391.

[2] T. van der Vorst, N. Jelicic, J. van Rees, R. N. Bekkers, R. Brennenraedts, R. Bakhyshov, Managing AI use in telecom infrastructures: Advice to the supervisory body on establishing risk-based AI supervision, Tech. rep., Dialogic innovatie & interactie (2020).

[3] A. Y. K. Al-Mulla, The Dark Side of AI in Telecom: Addressing Bias in Network Optimisation Models, International Journal of Computational Research and Development 10 (2) (2025) 34–42. doi:10.5281/zenodo.16081097.

[4] Trustpath.ai, From model drift to bias: 5 hidden AI risks every Telecom provider needs to address now, https://www.trustpath.ai/blog/from-model-drift-to-bias-5-hidden-ai-risks-every-telecom-provider-needs-to-address-now, accessed: 2025-08-10 (Mar. 2025).

[5] World Economic Forum, Artificial Intelligence in Telecommunications: White Paper, White paper, World Economic Forum, accessed: 2025-08-09 (Feb. 2025).

[6] A. Avinash, N. Manisha, Advancing Trustworthy AI for Sustainable Development: Recommendations for Standardising AI Incident Reporting, in: 2024 ITU Kaleidoscope: Innovation and Digital Transformation for a Sustainable World (ITU K), 2024, pp. 1–8. doi:10.23919/ITUK62727.2024.10772925.

[7] S. Avin, H. Belfield, M. Brundage, G. Krueger, J. Wang, A. Weller, M. Anderljung, I. Krawczuk, D. Krueger, J. Lebensold, et al., Filling gaps in trustworthy development of AI, Science 374 (6573) (2021) 1327–1329.

[8] O-RAN Alliance Next Generation Research Group (nGRG), Principles and Methodologies for AI/ML Testing in Next Generation Networks, Research Report RR-2024-05, O-RAN Alliance (May 2024).
URL https://mediastorage.o-ran.org/ngrg-rr/nGRG-RR-2 024-05-Principals\%20Methodologies\%20AIML\%20testing\%20next\%20Generation\%20Networks-v1.9.pdf

[9] Ericsson, Trustworthy ai - what it means for telecom, Tech. rep., Ericsson White Paper (June 2023).URL https://www.ericsson.com/en/reports-and-papers/white-papers/trustworthy-ai

[10] European Parliament, Artificial Intelligence Act (Regulation (EU) 2024/1689), https://eur-lex.europa.eu/legal-content/EN/TXT/?uri=CELEX%3A32024R1689 (2024).

[11] World Bank, Population, total, https://data.worldbank.org/indicator/SP.POP.TOTL, accessed: 2025-08-09 (2025).

[12] World Bank, Mobile cellular subscriptions, https://data.worldbank.org/indicator/IT.CEL.SETS, accessed: 2025-08-09 (2025).

[13] A. Johan, India's 5G Driving Improved Consumer Experience as Adoption Increases, https://www.ookla.com/articles/india-5g-experience-q4-2023, published: March 31, 2024; Accessed: 2025-08-09 (2024).

[14] Press Information Bureau, India, Expansion of 5G Network in the Country, https://www.pib.gov.in/PressReleasePage.aspx?PRID=2113855, posted: March 21, 2025; Accessed: 2025-08-09 (2025).

[15] Parliament of India, The Telecommunications Act, 2023, https://egazette.gov.in/WriteReadData/2023/250880.pdf, no. 44 of 2023 (2023).

[16] Parliament of India, The Digital Personal Data Protection Act, 2023, https://www.meity.gov.in/static/uploads/2024/06/2bf1f0e9f04e6fb4f8fef35e82c42aa5.pdf, act No. 22 of 2023 (2023).

[17] A. Mentges, L. Halekotte, M. Schneider, T. Demmer, D. Lichte, A resilience glossary shaped by context: Reviewing resilience-related terms for critical infrastructures, International journal of disaster risk reduction 96 (2023) 103893.

[18] P. Jiang, J. Rowsell, S. Schmidt, Crisis-ready telecom: Global approaches to emergency management in telecommunications, Telecommunications Policy (2025) 102914.

[19] P. O. Shoetan, O. O. Amoo, E. S. Okafor, O. L. Olorunfemi, Synthesizing AI'S impact on cybersecurity in telecommunications: a conceptual framework, Computer Science & IT Research Journal 5 (3) (2024) 594–605.

[20] I. Aviv, U. Ferri, Russian-Ukraine armed conflict: Lessons learned on the digital ecosystem, International Journal of Critical Infrastructure Protection 43 (2023) 100637.

[21] A. Agarwal, M. J. Nene, Addressing ai risks in critical infrastructure: Formalising the ai incident reporting process, in: 2024 IEEE International Conference on Electronics, Computing and Communication Technologies (CONECCT), 2024, pp. 1–6. doi:10.1109/CONECCT62155.2024.10677312.

[22] L. Sambucci, E.-A. Paracshiv, The accelerated integration of artificial intelligence systems and its potential to expand the vulnerability of the critical infrastructure, Romanian Journal of Information Technology and Automatic Control 34 (3) (2024) 131–148.

[23] A. S. George, When trust fails: Examining systemic risk in the digital economy from the 2024 crowdstrike outage, Partners Universal Multidisciplinary Research Journal 1 (2) (2024) 134–152.

[24] I. Linkov, K. Stoddard, A. Strelzoff, S. E. Galaitsi, J. Keisler, B. D. Trump, A. Kott, P. Bielik, P. Tsankov, Toward MissionCritical AI: Interpretable, Actionable, and Resilient AI, in: 2023 15th International Conference on Cyber Conflict: Meeting Reality (CyCon), IEEE, 2023, pp. 181–197.

[25] G. Lupo, Risky artificial intelligence: The role of incidents in the path to AI regulation, Law, Technology and Humans 5 (1) (2023) 133–152.

[26] A. Agarwal, M. J. Nene, A five-layer framework for AI governance: integrating regulation, standards, and certification, Transforming Government: People, Process and Policy 19 (3) (2025) 535–555. doi:10.1108/TG-03-2025-0065. URL https://doi.org/10.1108/TG-03-2025-0065

[27] V. Turri, R. Dzombak, Why we need to know more: Exploring the state of AI incident documentation practices, in: Proceedings of the 2023 AAAI/ACM Conference on AI, Ethics, and Society, 2023, pp. 576–583.

[28] S. McGregor, Preventing repeated real world AI failures by cataloging incidents: The AI incident database, in: Proceedings of the AAAI Conference on Artificial Intelligence, Vol. 35, no. 17, 2021, pp. 15458–15463.

[29] International Telecommunication Union, Resolution 69 (Rev. Kigali, 2022): Facilitating the creation of national computer incident response teams, particularly for developing countries, and cooperation among them, World Telecommunication Development Conference, Kigali, 2022, accessed: 202503-29 (2022). URL https://www.itu.int/dms_pub/itu-d/opb/res/D-RES-D.69-2022-PDF-E.pdf

[30] International Telecommunication Union, Resolution 50 (Rev. New Delhi, 2024): Cybersecurity, World Telecommunication Standardization Assembly, New Delhi, 15-24 October 2024, accessed: 2025-03-29 (2024). URL https://www.itu.int/dms_pub/itu-t/opb/res/T-RES-T.50-2024-PDF-E.pdf

[31] International Telecommunication Union, Resolution 58 (Rev. New Delhi, 2024): Encouraging the creation and enhancement of national computer incident response teams, particularly for developing countries, World Telecommunication Standardization Assembly, New Delhi, 15-24 October 2024, accessed: 2025-03-29 (2024). URL https://www.itu.int/dms_pub/itu-t/opb/res/T-RES-T.58-2024-PDF-E.pdf

[32] D. Tanevski, T. Janevski, V. Kafedziski, 225. survey on technical, business and regulatory aspects of artificial intelligence in telecom networks and services, Journal of Electrical Engineering and Information Technologies (2024) 108–115.

[33] M. Fenwick, E. P. Vermeulen, M. Corrales, Business and regulatory responses to artificial intelligence: Dynamic regulation, innovation ecosystems and the strategic management of disruptive technology, Robotics, AI and the Future of Law (2018) 81–103.

[34] AIID, AI Incident Database, https://incidentdatabase.ai, accessed: 2025-03-30 (2025).

[35] AIAAIC, AIAAIC Repository, https://www.aiaaic.org/aiaaic-repository, accessed: 2025-03-30 (2025).

[36] A. Agarwal, M. J. Nene, Standardised Schema and Taxonomy for AI Incident Databases in Critical Digital Infrastructure, in: 2024 IEEE Pune Section International Conference (PuneCon), 2024, pp. 1–6. doi:10.1109/PuneCon63413.2024.10895867.

[37] TelcoSec News, The Risks of AI in Life and Telecom: Lessons from T-Mobile's 2022 Incident, https://www.p1sec.com/blog/the-risks-of-ai-in-life-and-telecom-lessons-from-t-mobiles-2022-incident, published: December 31, 2024; Accessed: 2025-08-09 (2024).

[38] OECD, Defining AI Incidents and Related Terms, OECD Artificial Intelligence Papers 16, OECD Publishing, Paris (2024). doi:10.1787/d1a8d965-en.

[39] European Parliament, EU Artificial Intelligence Act Article 3: Definitions, https://artificialintelligenceact.eu/article/3/, accessed: 2025-03-30 (2024).

[40] O'Neil, Cathy, Weapons of Math Destruction: How Big Data Increases Inequality and Threatens Democracy, Crown, 2017.

[41] Parliament of India, The Information Technology (Amendment) Act, 2008, https://www.indiacode.nic.in/bitstream/123456789/15386/1/it_amendment_act2008.pdf, no. 10 of 2009 (2009).

[42] Ministry of Electronics and Information Technology, Government of India, Information Technology (The Indian Computer Emergency Response Team and Manner of Performing Functions and Duties) Rules, 2013, https://www.cert-in.org.in (2013).

[43] Ministry of Communications, Government of India, Telecommunications (Telecom Cyber Security) Rules, 2024, https://dot.gov.in/sites/default/files/Telecommunications%20%28Telecom%20Cyber%20Security%29%20Rules%2C%202024.pdf (2024).

[44] Ministry of Communications, Government of India, Telecommunications (Critical Telecommunication Infrastructure) Rules, 2024, https://dot.gov.in/sites/default/files/Telecommunications%20%28Critial%20Telecommunication%20Infrastructure%29%20Rules%2C%202024.pdf (2024).

[45] Ministry of Electronics and Information Technology, Government of India, Draft Digital Personal Data Protection Rules, 2025, https://www.meity.gov.in/static/uploads/2025/02/f8a8e97a91091543fe19139cac7514a1.pdf, released for public consultation, January 2025 (2025).

[46] Communications, Security, Reliability and Interoperability Council, WORKING GROUP 5: CYBER SECURITY INFORMATION SHARING Report on Information Sharing Barriers, Tech. rep., Federal Communications Commission (FCC) (June 2016). URL https://transition.fcc.gov/bureaus/pshs/advisory/csric5/WG5_Info_Sharing_Report_062016.pdf

[47] A. Zibak, A. Simpson, Cyber threat information sharing: Perceived benefits and barriers, in: Proceedings of the 14th international conference on availability, reliability and security, 2019, pp. 1–9.

[48] P. Koepke, Cybersecurity information sharing incentives and barriers, Sloan School of Management at MIT University: Cambridge, MA, USA (2017).

[49] W. Alkalabi, L. Simpson, H. Morarji, Barriers and incentives to cybersecurity threat information sharing in developing countries: a case study of saudi arabia, in: Proceedings of the 2021 Australasian Computer Science Week Multiconference, 2021, pp. 1–8.

[50] J. R. Brubacher, G. S. Hunte, L. Hamilton, A. Taylor, Barriers to and incentives for safety event reporting in emergency departments, Healthc Q 14 (3) (2011) 57–65.

[51] M. M. M. Hamed, S. Konstantinidis, Barriers to incident reporting among nurses: a qualitative systematic review, Western journal of nursing research 44 (5) (2022) 506–523.

[52] S. Aljabari, Z. Kadhim, Common barriers to reporting medical errors, The Scientific World Journal 2021 (1) (2021) 6494889.

[53] OECD, Oecd ai incidents monitor, https://oecd.ai/en/incidents-methodology, accessed: 2025-03-30 (2025).

[54] AVID, Ai vulnerability database, https://avidml.org/, accessed: 2025-03-30 (2025).

[55] European Parliament, Artificial Intelligence Act Section 2: Requirements for High-Risk AI Systems, https://artificialintelligenceact.eu/section/3-2/, accessed: 2025-03-30 (2024).

[56] Telecom Regulatory Authority of India, Recommendations on Leveraging Artificial Intelligence and Big Data in Telecommunication Sector, https://www.trai.gov.in/sites/default/files/2024-09/Recommendation_20072023.pdf (July 2023).

[57] Telecommunication Engineering Centre, Procedure for Mandatory Testing & Certification of Telecommunication Equipment. Version 3.0, Tech. Rep. TEC 93009:2024, Telecommunication Engineering Centre, New Delhi, India, Government of India (2024).

[58] A. Agarwal, M. Kumar, M. J. Nene, Enhancements for developing a comprehensive ai fairness assessment standard, in: 2025 17th International Conference on COMmunication Systems and NETworks (COMSNETS), 2025, pp. 1216–1220. doi:10.1109/COMSNETS63942.2025.10885551.